\documentclass[twocolumn,showpacs,amsmath,amstex,amssymb,mathfonts]{revtex4}
\usepackage{graphicx,epstopdf,verbatim,color,ulem}

\begin{document}
\title{Common path interference in Zener tunneling is a universal phenomenon}
\author{Sonika Johri$^{1}$, Rahul Nandkishore$^{2}$,  R. N. Bhatt$^{1}$ and E.J.Mele$^{3}$}
\affiliation{$^1$ Department of Electrical Engineering, Princeton University, Princeton, NJ 08544}
\affiliation{$^2$ Princeton Center for Theoretical Science, Princeton University, Princeton, NJ 08544}
\affiliation{$^3$ Department of Physics and Astronomy, University of Pennsylvania, Philadelphia, PA 19104, USA}

\begin{abstract}
We show that the probability of electric field induced interband  tunneling in solid state systems is generically a non-monotonic (oscillatory) function of the applied field. This unexpected behavior can be understood as arising due to a common path interference between two distinct tunneling solutions. The phenomenon is insensitive to magnetic field, and arises whenever the low energy dispersion relation contains higher order terms in addition to the usual $p^2$ term. Such higher order terms are generically present, albeit with small co-efficient, so that the oscillatory Zener tunneling is a universal phenomenon. However, the first `Zener oscillation' occurs at a transmission probability which is exponentially small when the co-efficient of the higher order terms is small. This explains why this oscillatory aspect of Zener tunneling has been hitherto overlooked, despite its universality. The common path interference is also destroyed by the presence of odd powers of $p$ in the low energy dispersion relation. Since odd powers of $p$ are strictly absent only when the tunneling barrier lies along an axis of mirror symmetry, it follows that the robustness of the oscillatory behavior depends on the orientation of the tunneling barrier. Bilayer graphene is identified as a particularly good material for observation of common path interference, due to its unusual nearly isotropic dispersion relation, where the $p^4$ term makes the leading contribution.  
\end{abstract}


\maketitle

\section{Introduction}
Tunneling and interference are two signature quantum effects that are commonly thought to occur in very different experimental setups. Interference usually involves the superposition of distinct saddle point solutions, whereas tunneling through a barrier proceeds through the evanescent solution with maximum decay length. Recently, it was shown that particles in bilayer graphene exhibit interference underneath a tunnel barrier, leading to a Zener tunneling probability that is a non-monotonic function of barrier height \cite{rahul}. The interference was further identified to be of a common path type, and hence unaffected by magnetic fields. The tunneling intensity is an oscillatory function of the bandgap and the potential:
\begin{equation}
T=4|a|^2 e^{\frac{-2}{\hbar}S_i}\cos^2\left(\frac{S_r}{\hbar}+\phi\right)
\label{eq:T}
\end{equation}
where $S_r$ and $S_i$ are monotonic functions of the bandgap and field strength, and $a$ and $\phi$ are constants. The bandgap and field strength can be tuned so that the tunneling current becomes zero.

In this work we show that such common path interference underneath the tunnel barrier is not specific to bilayer graphene, but is in fact generic, occurring in any system where the dispersion relation contains a $p^4$ term in addition to the usual $p^2$ term. Since higher derivative terms generically arise in effective low energy Hamiltonians \cite{narrowgap}, the common path interference is universal. However, if the coefficient of the high derivative term is small, then the first `Zener oscillation' does not manifest itself until an exponentially small tunneling probability is reached, explaining why this common path interference effect has hitherto been overlooked. The unique feature of bilayer graphene is that the higher derivative terms do not come with small coefficients, and thus the Zener oscillation phenomenon should be (relatively) easy to see. We also show that there is nothing special about a $p^4$ term, and that common path interference generally arises when the dispersion relation contains higher powers of $p^2$ (but the odd powers of $p$ are absent). Common path interference may be detected either in transport experiments, or in optical absorption experiments designed to detect photon assisted tunneling. Bilayer (and ABC) multilayer graphene are identified as particularly promising candidates for detection of this phenomenon, but other narrow gap semiconductors (such as several recently discovered topological insulators) may also offer a potential experimental playground.


This paper is structured as follows. We begin in section II by explaining the origin of the Zener oscillations using the WKB approximation. We generalize the analysis in \cite{rahul} to a dispersion relation with a quadratic piece, and demonstrate that the common path interference survives. In section \ref{sec: generalsolution}, we show that in fact common path interference is a generic feature for dispersion relations that contain higher powers of the momentum. Finally, in section III, we verify our results (for the fourth order polynomial dispersion relation) by re-deriving them in an independent manner, by mapping the problem to a differential equation which can be numerically solved. Finally, in section IV, we present our conclusions and comment on the possibilities for observation in experiment.

\section{WKB Approximation}
In this section we calculate the probability of Zener tunneling as a function of applied field using the Wentzel-Kramers-Brillouin (WKB) approximation. A convenient toy model to investigate the physics we have in mind is the two band effective Hamiltonian for bilayer graphene \cite{mccann, CN}
\[ H = \left( \begin{array}{ccc}
\Delta & \frac{p^2}{2m}e^{i2\phi} \\
\frac{p^2}{2m}e^{-i2\phi} & -\Delta \end{array} \right)+\beta\left( \begin{array}{ccc}
p^2 & 0 \\
0 & -p^2 \end{array} \right). \label{eq: H}\] 
Here, $\Delta=E_g/2$, where $E_g$ is the band gap, which can be tuned using external gates. The work \cite{rahul} analyzed the above Hamiltonian at $\beta = 0$, however, a rigorous treatment of the effects of higher bands generates a small but nonzero $\beta<0$ \cite{mccann}. Note that $\beta < 0$ leads to bands that have a `mexican hat' shape, whereas $\beta>0$ leads to bands that have only one minimum. A great advantage of the Hamiltonian above is that the low energy dispersion has the form
\begin{equation}
|E| = \Delta + \sum_n a_n p^{2n} \label{eq: polynomial}
\end{equation}
which is the general polynomial dispersion of interest to us in this paper. A quadratic dispersion ($a_{n\ge2} = 0$) is conventional, and arises for e.g.free fermions governed by the Schrodinger equation. The tunneling behavior in this limit is well understood, and does not display common path interference. However, terms at higher order in the momentum are generically present in the dispersion relation about the bottom of any bandstructure (other than free fermions) as can be explicitly seen, e.g. by performing $\vec{k}.\vec{p}$ perturbation theory for order higher than two \cite{narrowgap}. This paper is about what happens when these higher momentum terms are nonzero. It is sufficient to include just the first non-trivial term, $a_2 \neq 0$. We will demonstrate that this leads to common path interference, and a tunneling probability that is a non-monotonic function of barrier height. Including further terms $a_3, a_4...$ does not change the physics, as we show in Section III. We note that an added advantage of the Hamiltonian (\ref{eq: H}) is that $a_1 \propto \beta$. Thus, by taking $\beta \rightarrow 0$, we can recover the model analyzed in \cite{rahul}, which serves as a useful double check on our results. 

We will consider the Hamiltonian (\ref{eq: H}) with a one-dimensional linear potential $V(x) = F x$, where $F$ is the force on an electron due to an (in-plane) electric field. The treatment here can be easily generalized to other monotonic potentials. In the presence of this potential, Zener tunneling of particles between conduction and valence bands will occur \cite{Zener}. We calculate the probability of Zener tunneling using a WKB approximation, in the manner of \cite{rahul}. 

For a slowly varying potential along the $x$-axis, the WKB approximation states that the change in the wavefunction after propagating from $x_0$ to $x$ is proportional to $e^{i\int_{x_0}^{x} p_x(x) dx}$. To find the amplitude of the tunneling through the barrier, we have to solve for $p_x$ and integrate it over the forbidden region of the energy gap, in which it has a non-zero imaginary component.

For simplicity let us consider just electrons at normal incidence, $p_y=0$. The inclusion of a non-zero $p_y$ complicates the analysis without changing the physics. We will focus on electrons at normal incidence in this paper. Then the momentum along the x-direction, $p_x = p$. Finding eigenvalues of $H+V(x)$ yields the equation :
\begin{eqnarray}
&&p^4+\frac{2\Delta\beta}{\frac{1}{4m^2}+\beta^2}p^2+\frac{\Delta^2-E'^2}{\frac{1}{4m^2}+\beta^2}=0
\end{eqnarray}
where $E'=E-V(x)$. Let us define for brevity the quantities
\begin{equation}
c=\frac{1}{4m^2}+\beta^2,\quad  \gamma=\frac{\Delta\beta}{c}, \quad \delta=\frac{\Delta^2-E'^2}{c}.
\end{equation}

Therefore,
$p^4+2\gamma p^2+\delta=0$.
The solutions to this quadratic equation in $p^2$ are:

\begin{equation}
p^2=-\gamma\pm\sqrt{\gamma^2-\delta}
\end{equation}

This equation implies that there are three distinct regions with sharply different physical properties (Fig. 1), and we will discuss each region in turn. The essential ideas are illustrated in Fig.1 (for $\beta < 0$) and Fig.2 (for $\beta >0$) respectively. 

\subsection{Region I: $\delta<0$}

 In this region there are 2 real solutions:
\begin{equation}
p_x=\pm\sqrt{-\gamma+\sqrt{\gamma^2-\delta}}
\end{equation}
and two imaginary ones:
\begin{equation}
p_x=\pm i\sqrt{\gamma+\sqrt{\gamma^2-\delta}}
\end{equation}
The former two solutions correspond to propagating waves moving left and right respectively, while the latter correspond to evanescent waves. This region corresponds to the region outside the tunneling barrier. 

\subsection{Region II: $0<\delta<\gamma^2$}

For $\beta<0$, this region is outside the tunneling barrier. However, there are four real solutions given by
\begin{equation}
p_x=\pm\sqrt{-\gamma\pm\sqrt{\gamma^2-\delta}}.
\end{equation}
To understand the existence of four solutions, we have only to remember that the bandstructure for $\beta < 0$ is a mexican hat. A line cut at low energy will intersect the mexican hat at four points, and these four points correspond to the four real solutions identified above. 

Meanwhile for $\beta>0$, there are 4 imaginary solutions given by
\begin{equation}
p_x=\pm i\sqrt{\gamma\pm\sqrt{\gamma^2-\delta}}
\end{equation}
Thus, for $\beta >0$, this region is inside the tunnel barrier. However, in this region the tunneling solutions are non-degenerate (have different decay lengths), and tunneling will be dominated by the solution with largest decay length. 

\subsection{Region III: $\delta>\gamma^2$}

In this region, the solutions are all complex, and take the form
\begin{eqnarray}
\nonumber p_x &=& \pm (-\gamma\pm i\sqrt{-\gamma^2+\delta})^{1/2}\\
&=&\pm \left( \sqrt{\frac{\sqrt{\delta}-\gamma}{2}} \pm i \sqrt{\frac{\sqrt{\delta}+\gamma}{2}}\right)
\end{eqnarray}
The lack of real solutions means that this region is inside the tunneling barrier. Only two of the solutions above correspond to tunneling from left to right. However, both are degenerate (have the same decay length). As a result, both will contribute equally to tunneling, giving rise to common path interference. 
\begin{figure}
\includegraphics[width=\columnwidth]{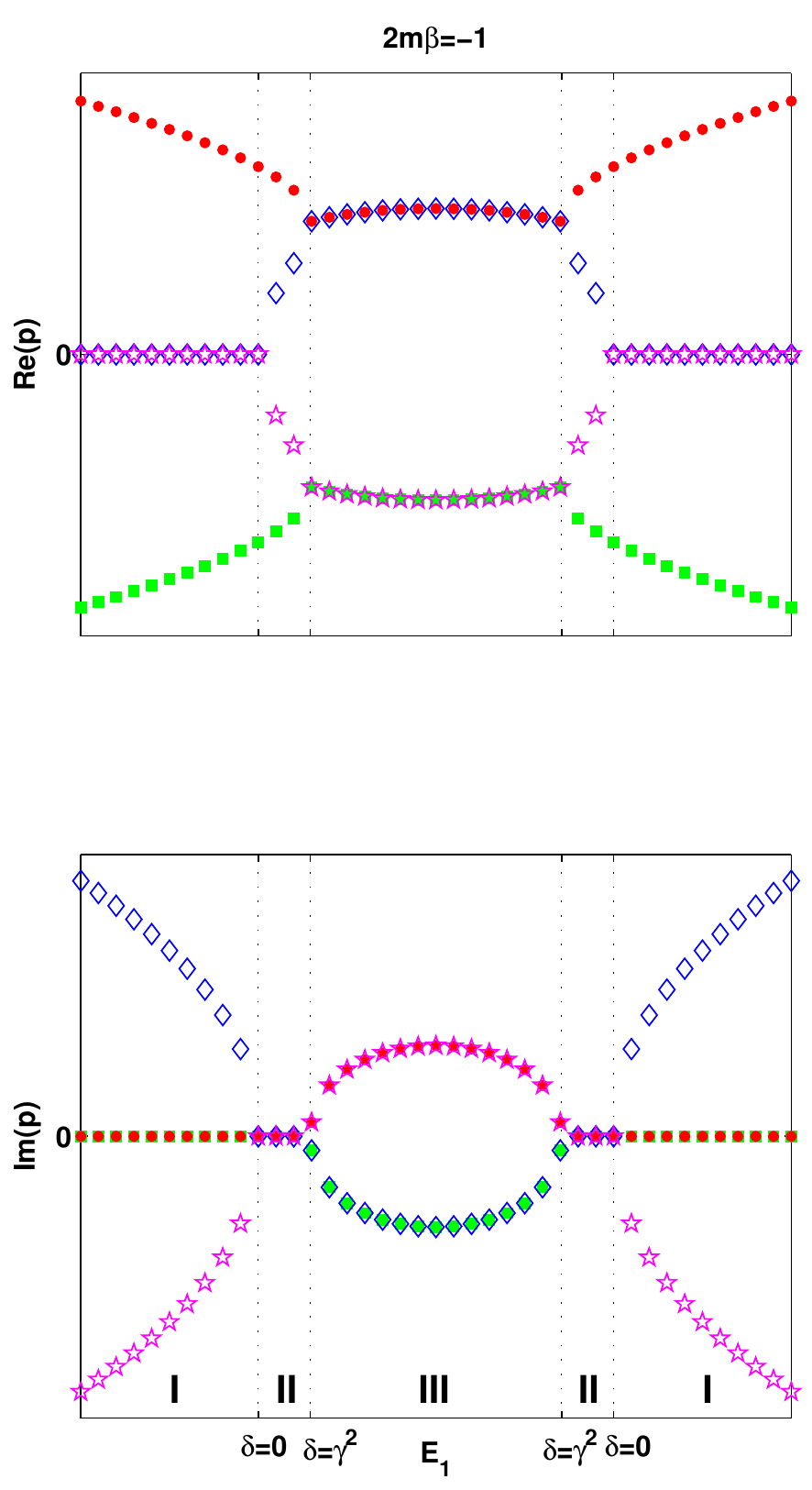}
\caption{Real (above) and imaginary (below) parts of $p_x$ at $2m\beta=-1$.Red circles are $\sqrt{-\gamma+\sqrt{\gamma^2-\delta}}$, green squares are -$\sqrt{-\gamma+\sqrt{\gamma^2-\delta}}$, blue diamonds are $\sqrt{-\gamma-\sqrt{\gamma^2-\delta}}$, and purple stars are $-\sqrt{-\gamma-\sqrt{\gamma^2-\delta}}$.}
\label{fig:a_-1}
\end{figure}
\begin{figure}
\includegraphics[width=\columnwidth]{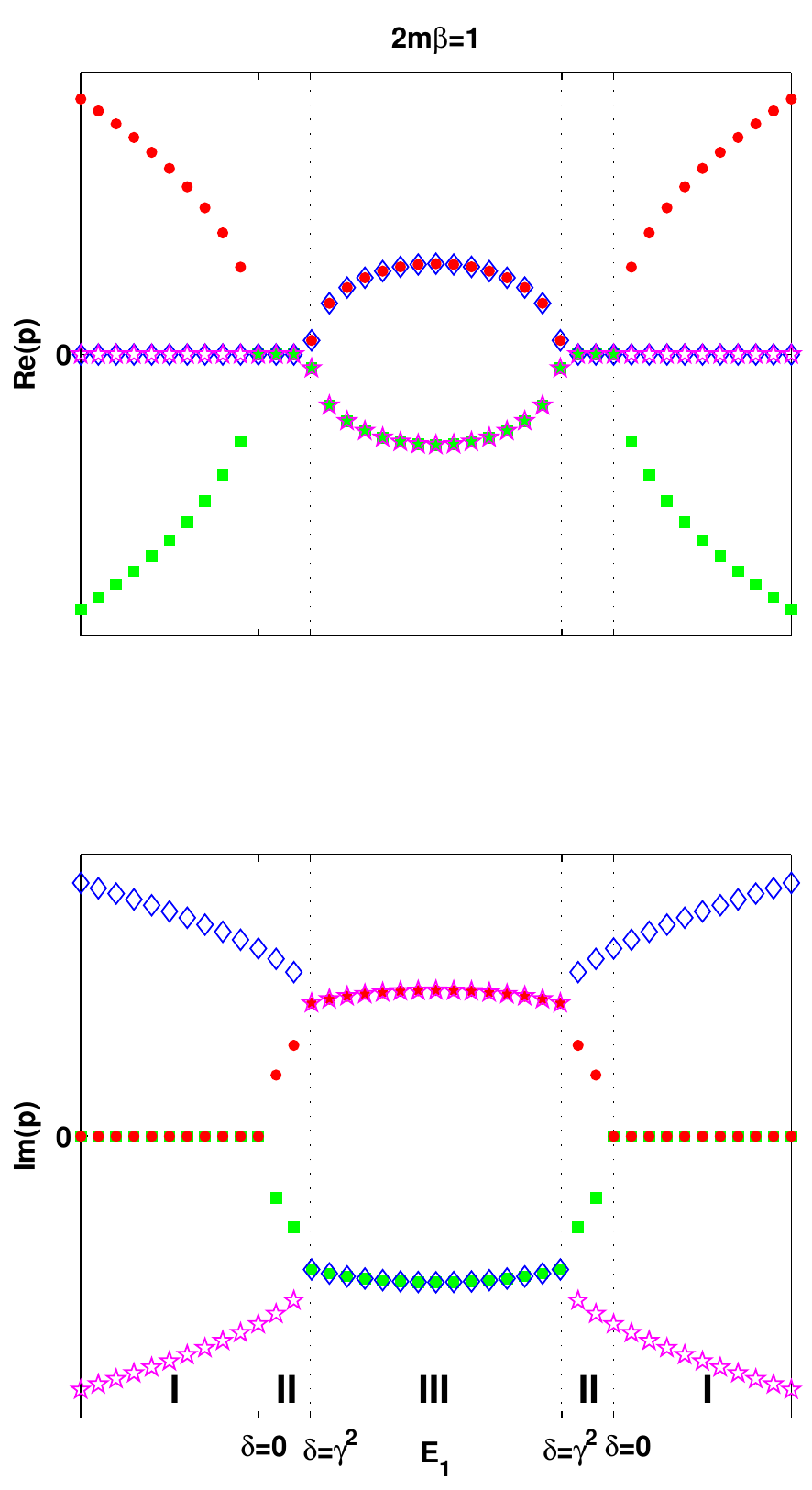}
\caption{Real (above) and imaginary (below) parts of $p_x$ at $2m\beta=1$. Red circles are $\sqrt{-\gamma+\sqrt{\gamma^2-\delta}}$, green squares are -$\sqrt{-\gamma+\sqrt{\gamma^2-\delta}}$, blue diamonds are $\sqrt{-\gamma-\sqrt{\gamma^2-\delta}}$, and purple stars are $-\sqrt{-\gamma-\sqrt{\gamma^2-\delta}}$.}
\label{fig:a_1}
\end{figure}
\subsection{WKB solution}
The real and imaginary parts of $p$ are depicted pictorially in Figs. \ref{fig:a_-1} and \ref{fig:a_1} for a negative and positive value of $\beta$ respectively. For tunneling from left to right in these figures, we choose solutions so that $p_x$ is always continuous. Also, in regions II and III, any imaginary component of $p_x$ should have positive sign, so that that the solution decays to the right. Under these constraints, the possible solutions for $\beta < 0$ (the relevant case for BLG) reduce to\\
Region I: $p_x=\pm\sqrt{-\gamma+\sqrt{\gamma^2-\delta}}$\\
Region II: $p_x=\pm\sqrt{-\gamma+\sqrt{\gamma^2-\delta}}$\\
Region III: $p_x=\pm  \sqrt{\frac{\sqrt{\delta}-\gamma}{2}} + i \sqrt{\frac{\sqrt{\delta}+\gamma}{2}}$\\
Note that in region III there are two degenerate tunneling solutions. We will show that interference of these solutions will lead to an expression of the form (\ref{eq:T}). Meanwhile, 
for $\beta>0$ we obtain the solutions\\
Region I: $p_x=\pm\sqrt{-\gamma+\sqrt{\gamma^2-\delta}}$\\
Region II: $p_x=i\sqrt{-\gamma\pm\sqrt{\gamma^2-\delta}}$\\
Region III: $p_x=\pm  \sqrt{\frac{\sqrt{\delta}-\gamma}{2}} + i \sqrt{\frac{\sqrt{\delta}+\gamma}{2}}$\\
Again there are degenerate tunneling solutions in region III, which will give rise to common path interference. 

We now explicitly evaluate the tunneling probability. 
The end-points $x_L$ and $x_R$ of the barrier region are the points at which $\gamma^2 = \delta$. This yields:
\begin{eqnarray}
x_{L,R}&=&\left(-E\pm\frac{\Delta}{2m\sqrt{c}}\right)\frac{1}{F}
\end{eqnarray}
Note that here any form of $V(x)$ can be used to obtain the endpoints, as long as it is smooth and monotonic.

For $\beta<0$, the barrier is only in region III. The change in the wavefunction after it has moved through region III is proportional to $e^{I_{3b}}(e^{iI_{3a}}+e^{-iI_{3a}})$, where $I_{3a}$ and $I_{3b}$ are the integrals over the barrier region of the real and imaginary part of the momentum respectively. We now explicitly evaluate these integrals. We have 
\begin{eqnarray}
\nonumber I_{3a}&=&\frac{1}{\sqrt{2}\hbar}\int_{x_L}^{x_R}\sqrt{\sqrt{\delta}-\gamma}dx\\
&=&\frac{1}{\sqrt{2}\hbar}\int_{x_L}^{x_R}\sqrt{\sqrt{\frac{\Delta^2-E'^2}{c}}-\frac{\Delta\beta}{c}}dx
\end{eqnarray}
Changing variables to $E'=E+Fx$, we obtain 
\begin{eqnarray}
\nonumber I_{3a}&=&\frac{1}{Fc^{1/4}\sqrt{2}\hbar}\int_{E'_L}^{E'_R}\sqrt{\sqrt{\Delta^2-E'^2}-\frac{\Delta\beta}{\sqrt{c}}}dE'\\
\end{eqnarray}
where $E'_L=E+Fx_L=\frac{-\Delta}{2m\sqrt{c}}$ and $E'_R=E+Fx_R=\frac{\Delta}{2m\sqrt{c}}$.
After scaling out all dimensionful quantities, this can be recast as 
\begin{eqnarray}
\nonumber I_{3a}&=&\frac{2\Delta^{3/2}m^{1/2}}{F\hbar(1+a^2)^{1/4}}\int_{0}^{\theta_0}\sqrt{\cos(\theta)+\cos(\theta_0)}cos(\theta)d\theta\\
&=&\left(\frac{\Delta}{\Delta_0}\right)^{3/2}\alpha_{3a}(2m\beta)
\end{eqnarray}
where $\Delta_0=\left(\frac{(F\hbar)^2}{2m}\right)^{1/3}$ and $\theta_0=\cos^{-1}(|2m\beta|/\sqrt{1+(2m\beta)^2})$ are the two fixed parameters that characterize a particular tunneling process. The function $\alpha_{3a}$ is plotted in Fig.3. 

Similarly, $I_{3b}$ is given by
\begin{eqnarray}
\nonumber I_{3b}&=&\frac{2\Delta^{3/2}m^{1/2}}{F(1+a^2)^{1/4}}\int_{0}^{\theta_0}\sqrt{\cos(\theta)-\cos(\theta_0)}\cos(\theta)d\theta\\
&=&\left(\frac{\Delta}{\Delta_0}\right)^{3/2}\alpha_{3b}(2m\beta)
\end{eqnarray}
Again, $\alpha_{3b}$ is plotted in Fig.3b. Note that $\alpha_{3a}$ and $\alpha_{3b}$ are both strictly positive. 

Meanwhile, for $\beta>0$, for the tunneling through region III will have exactly the same form as for $\beta < 0$, except that $\alpha_{3a}$ and $\alpha_{3b}$ will interchange with each other, i.e. $\alpha_{3a}(\beta > 0) = \alpha_{3b} (\beta < 0)$ and $\alpha_{3b}(\beta > 0) = \alpha_{3a} (\beta < 0)$. Also there will be a contribution to the decay of the wavefunction amplitude as it moves through region II which is of the form
\begin{eqnarray}
\nonumber I_2&=&\frac{2^{1/2}\Delta^{3/2}m^{1/2}}{\hbar F(1+a^2)}\int_{\theta_0}^{\Pi/2}\sqrt{\cot(\theta)-\cot(\theta_0)}\cot(\theta)\csc(\theta)d\theta\\
&=&\left(\frac{\Delta}{\Delta_0}\right)^{3/2}\alpha_2(2m\beta)
\end{eqnarray}
where $\alpha_{2}$ is plotted in Fig.3. After tunneling, the wavefunction will be multiplied by a factor $e^{-(I_{3a}+2I_2)}(e^{iI_{3b}}+e^{-iI_{3b}})$. Note that once again we will have common path interference coming from region III, which will give rise to an oscillatory tunneling probability (\ref{eq:T}). 
\begin{figure}
\includegraphics[width=\columnwidth]{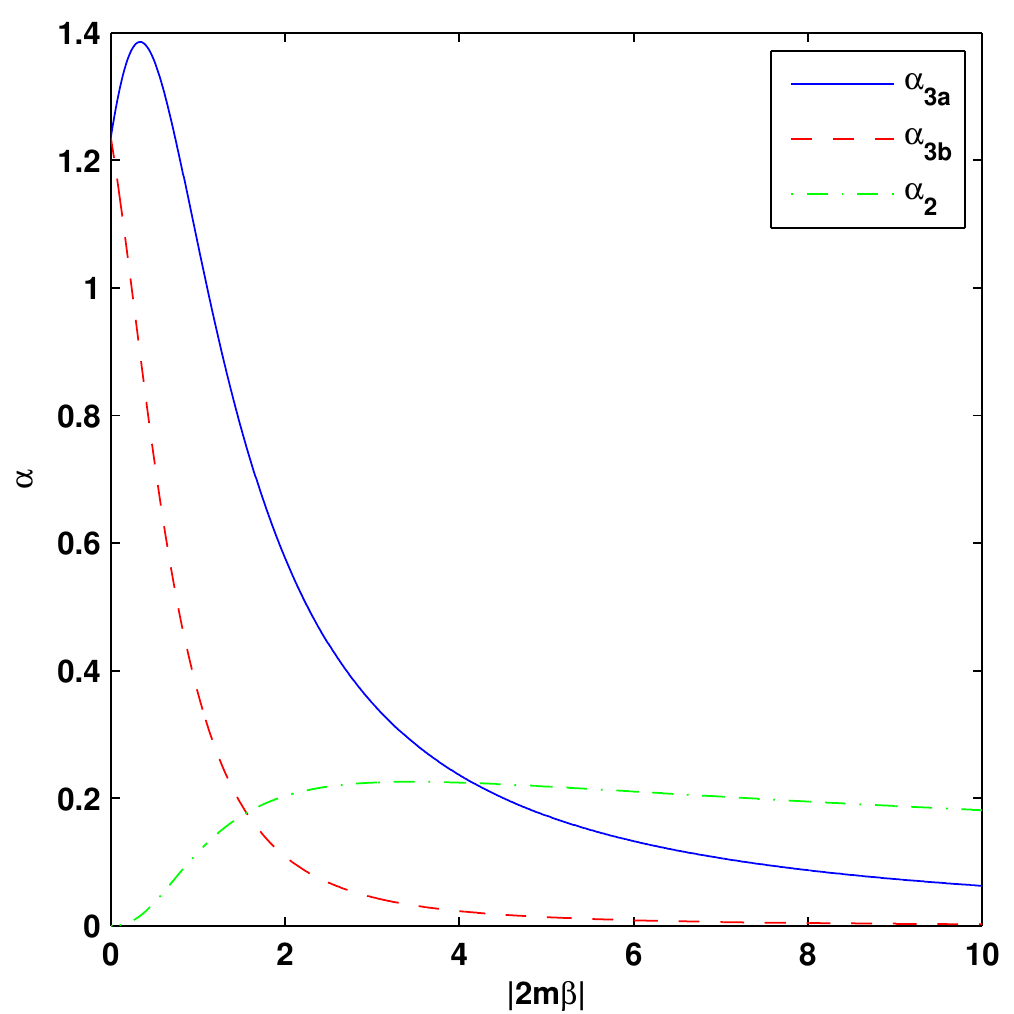}
\caption{$\alpha_1$, $\alpha_2$ and $\alpha_3$ vs $|2m\beta|$.}
\label{fig:alpha}
\end{figure}

Thus, we have shown that common path interference occurs and results in tunneling oscillations whenever the dispersion relation has a quartic component. The existence of common path interference does not depend on the sign of the quadratic component. However, as the ratio of quadratic to quartic co-efficients is made bigger, the tunneling oscillations are pushed to larger and larger bandgap, and the first antinode occurs at exponentially smaller tunneling amplitude (Fig.4). This follows because the exponential suppression of tunneling is proportional to $\alpha_3$, and $\alpha_3$ goes to zero as $2m\beta \rightarrow \infty$. 

In most materials, the co-efficient of the quartic term is much smaller than the corresponding coefficient of the quadratic term. As a result, the Zener oscillations, while theoretically present, will be extremely weak, and may not be visible experimentally. The special feature of bilayer graphene is that $2m \beta \ll 1$, and as a result the Zener oscillations should be relatively easy to see. 

\section{Generality of common path interference}
\label{sec: generalsolution}
In this section we demonstrate that common path interference is not simply a property of quartic dispersions, but in fact is a general feature for any dispersion of the form
\begin{equation}
|E| = \Delta + \sum_{n=1}^{N} a_n p^{2n} \label{eq: generalsolution}
\end{equation}
with real coefficients $a_n$. The argument is simple and proceeds as follows. 

The equation (\ref{eq: generalsolution}) is a $N^{th}$ order polynomial in $p^2$ with real co-efficients. By the complex conjugate root theorem, complex solutions must come in conjugate pairs. Additionally, in the barrier region, there cannot be a real positive solution for $p^2$. This implies that for even $N$, the solutions to (\ref{eq: generalsolution}) must have the form $p^2 = \{r_j e^{\pm i 2 \phi_j}\}$, where $j = 1...N/2$ and $r$ and $\phi$ are real numbers. The corresponding solutions for $p$ are $p = \{\sqrt{r_j} e^{\pm i \phi_j}, \sqrt{r_j} e^{\pm i \phi_j + \pi}\}$. The solutions corresponding to tunneling from left to right are $p = \{ \sqrt{r_j} (\pm \cos \phi_j + i \sin \phi_j)\}$, which all come in degenerate pairs. The system simply picks out the degenerate pair with longest penetration length (minimum $\sqrt{r_j} \sin \phi_j$). Since there are two solutions with this decay length, the two solutions can interfere, to give rise to the physics discussed above. Note that this is true for any (even order) polynomial dispersion relation, not just the quartic dispersion relation we discussed thus far. 

For odd $N$ the story is similar, except that there is also one solitary solution $p^2 = - A^2$, where A is a real number. This leads to solutions $p = \pm i A$. Common path interference can be absent for odd $N$ if and only if $A$ is the longest penetration length in the problem i.e. IFF $A < \min{\sqrt{r_j} \sin \phi_j}$. If this very special condition is satisfied then there will be no common path interference, and this single non-degenerate mode will dominate tunneling. Otherwise the story will be similar to that for even $N$, and there will be common path interference. 

Thus, we have shown that for a dispersion relation that is an $N^{th}$ order polynomial in $p^2$ (with real co-efficients), there must necessarily be common path interference if $N$ is even. If $N$ is odd the system may escape common path interference, but only if the single non-degenerate tunneling mode has longer decay length than the $(N-1)$ degenerate tunneling modes. 

We have focused on electrons at normal incidence, since tunneling at oblique angles is exponentially supressed by the longer tunneling path. We have also assumed that the dispersion relation does not contain odd powers of $p$ (i.e. the dispersion relation is symmetric under $p \rightarrow -p$). This is rigorously true for a tunneling barrier parallel to an axis of reflection symmetry of the underlying crystal. 

For a general tunneling barrier, however, odd powers of $p$ will arise in the dispersion relation due to crystal fields, albeit with small co-efficient. These odd powers will lift the degeneracy of the tunneling solutions, and will ultimately destroy common path interference. However, if the co-efficients of the odd $p$ terms are small, then the degeneracy will be barely lifted, and the degeneracy lifting will only manifest itself on lengthscales much larger than the tunneling decay length. In this event, common path interference should still be visible if the barrier is no too wide - i.e. we should still be able to see the first few oscillations, but high order nodes will be washed out by the lifting of the degeneracy of the two interfering modes. 

\section{Solution by mapping to differential equation}
\begin{figure}
\includegraphics[width=\columnwidth]{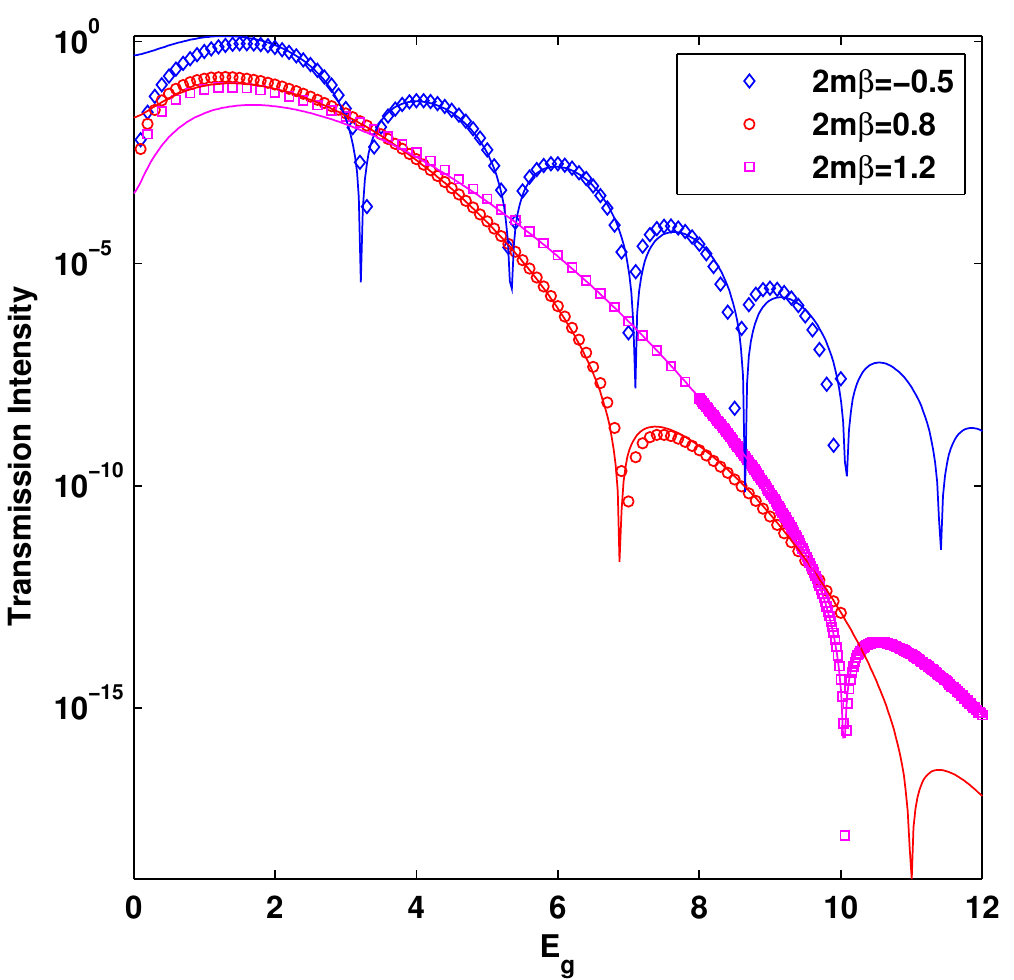}
\caption{Transmission as a function of the bandgap for different values of $2m\beta$, $E_g$ measured in units of $\Delta_0$. The points are numerical data and the lines are fits to the points with functions of the form \ref{eq:T}. $S_r$ and $S_i$ are obtained from the WKB approximation while $a$ and $\phi$ are chosen for the best fit.}
\label{fig:trans}
\end{figure}
An independent way to verify the conclusions from the WKB treatment in real space is to consider the problem in momentum space. Following Ref. \cite{rahul}, we substitute $x=i\hbar\partial p_x$. This allows us to write the Schrodinger equation $H \Psi = E \psi$ as 
\begin{equation}
i\hbar F \frac{\partial \Psi}{\partial p_x}=\left(\frac{p_x^2-p_y^2}{2m}\sigma_1 -\frac{2p_xp_y}{2m}\sigma_2+(\Delta+\beta p_x^2)\sigma_3\right)\Psi
\end{equation}
where the $\sigma_i$ are the Pauli matrices, and we have taken $E=0$ by choosing the origin of coordinates appropriately. This equation can be interpreted as a two state system swept through an avoided crossing, with $p_x$ playing the role of time. The probability of Zener tunneling is equal to the probability that a system initialized in a particular eigenstate ends up in the orthogonal eigenstate after the sweep through the avoided crossing. Again, we focus on electrons at normal incidence, $p_y = 0$. 

For $|2m\beta|<1$, as $p_x=\pm \infty$, the eigenstates of the Hamiltonian are eigenstates of $\sigma_1$, i.e. $\phi_1=[1, 1]$ and $\phi_2=[1, -1]$. The Zener tunneling probability is equal to the probability of starting in the state $\phi_1$ and ending in the state $\phi_2$ after the sweep through the avoided crossing. In a simple approximation, we can argue that  the transition has a chance of happening when the coefficient of the $\sigma_3$ term is larger than that of the $\sigma_1$ term: 
\begin{eqnarray}
\nonumber \frac{p_x^2}{2m}&<&\Delta+\beta p_x^2\\
|p_x|&<&\sqrt{\frac{2m\Delta}{1-2m\beta}}=p_{\Delta}
\end{eqnarray}
Let the wavefunction during the transtion be $\Psi=C_1\phi_1+C_2\phi_2$. $|C_1|^2+|C_2|^2=1$. This gives us the two simultaneous differential equations:
\begin{eqnarray}
i\hbar F \frac{dC_1}{dp_x}&=&(\Delta+\beta p_x^2)C_2\\
i\hbar F \frac{dC_2}{dp_x}&=&(\Delta+\beta p_x^2)C_1
\end{eqnarray}
with the initial conditions at $p_x=-p_{\Delta}$, $|C_2|=1$ and $|C_1|=0$.

These are solved by
\begin{eqnarray}
C_1&=&\sin\left(\omega\left(p_x+\frac{\beta}{3\Delta}p_x^3\right)+A\right)\\
C_2&=&i \cos\left(\omega\left(p_x+\frac{\beta}{3\Delta}p_x^3\right)+A\right)
\end{eqnarray}
where, $\omega=\frac{\Delta}{\hbar F}$.
From initial conditions, $A=\omega p_{\Delta}\left(1+\frac{\beta}{3\Delta}p_{\Delta}^2\right)$.

The total phase acquired at $p_x=p_{\Delta}$ is
\begin{equation}
\theta=2\omega p_{\Delta}\left(1+\frac{\beta}{3\Delta} p_{\Delta}^2\right)=\left(\frac{\Delta}{\Delta_0}\right)^{3/2}\alpha(m\beta)
\end{equation}
which indicates that the probability of tunneling ($|C_2|^2$) will oscillate with period dependent on $m\beta$.

For $|2m\beta|>1$ however, the coefficient of the $\sigma_3$ term is always larger than that of the $\sigma_1$ term. 
The simple approximation above is no longer applicable then since at no time does the transition term ($\sigma_1$) in this case become larger than the other term. The transition may still happen but the complete differential equation has to be considered.

Since an analytic approximation is possible for only a limited range of values of $\beta$ in this formulation of the problem, the differential equation was also solved numerically. We evolve the wavefunction starting from $p_x=-30\sqrt{2m\Delta}$ to $p_x=30\sqrt{2m\Delta}$ and the results for the tunneling so obtained are shown in Fig. \ref{fig:trans} for three representative cases of $\beta<0$, $\beta>0$ and $\beta>1/(2m)$ are shown. They show clearly that the oscillations occur even outside the limit $|2m\beta|<1$. Note too that the numerical data (points in Fig.4) is in good agreement with the WKB result (lines in Fig.4), except at very small values of $E_g$. The deviation of the numerical solution from the WKB estimate at small values of $E_g$ (in units of $\Delta_0$) are to be expected since this is precisely the limit where the WKB approximation fails. 

Note too that as $|\beta|$ is made bigger, the first oscillation is pushed to ever higher values of $E_g$, and the transmission probability at the first antinode becomes exponentially smaller. In generic systems, where $\beta \gg 1$ (i.e. the quadratic term in the dispersion dominates over the quartic term), the transmission probability at the first non-trivial antinode will be extremely small, and will likely not be detectable given limited experimental resolution. This explains why the common path interference phenomenon has so far been overlooked, despite its universality. However, in bilayer graphene, $\beta \ll 1$ and observation of the common path interference may be a realistic possibility. 

\section{Outlook}
We have shown that common path interference occurs generically when the low-order energy dispersion relation contains higher order terms in addition to the usual $p^2$ term. When the higher order terms come with small co-efficients, then the Zener oscillations get pushed to very large values of the tunneling barrier, such that the first antinode involves an exponentially small transmission probability. In this situation Zener oscillations may not be visible. However, in bilayer (or ABC multilayer \cite{Zhang}) graphene, the leading term in the dispersion relation is a $p^4$ term ($p^{2N}$) term. As a result, in the bilayer or ABC multilayer, the Zener oscillation effect manifests itself at uniquely small barrier heights, and the first antinode has unusually high transmission probability. Thus, bilayer or ABC multilayer graphene is the ideal place to look for oscillations in Zener transmission. 

Other materials with relatively large corrections to parabolic spectrum include narrow-gap semiconductors, including the recently discovered topological insulators. Typical narrow-gap materials are HgCdTe alloys and InSb. They are popular infrared photo-detector materials. The Kane model is commonly used to describe the bands near the $\Gamma$ point when the bands are non-parabolic. The generic form of bands in this model is $E=a+bp^2+\sqrt{c+dp^2}$, where $a$, $b$, $c$ and $d$ are constants which can be zero. These bands will always satisfy the constraints for common path interference. 

In this paper we considered only the tunneling of electrons at normal incidence. The full junction $IV$ characteristic may be obtained by integrating over angles of incidence, in the manner of \cite{rahul}. The resulting $IV$ curve are expected to contain regions of negative differential resistance, as pointed out in \cite{rahul}, with the negative differential resistance being a consequence of common path interference. This may be useful from a device physics standpoint. 

\begin{figure}
\includegraphics[width=\columnwidth]{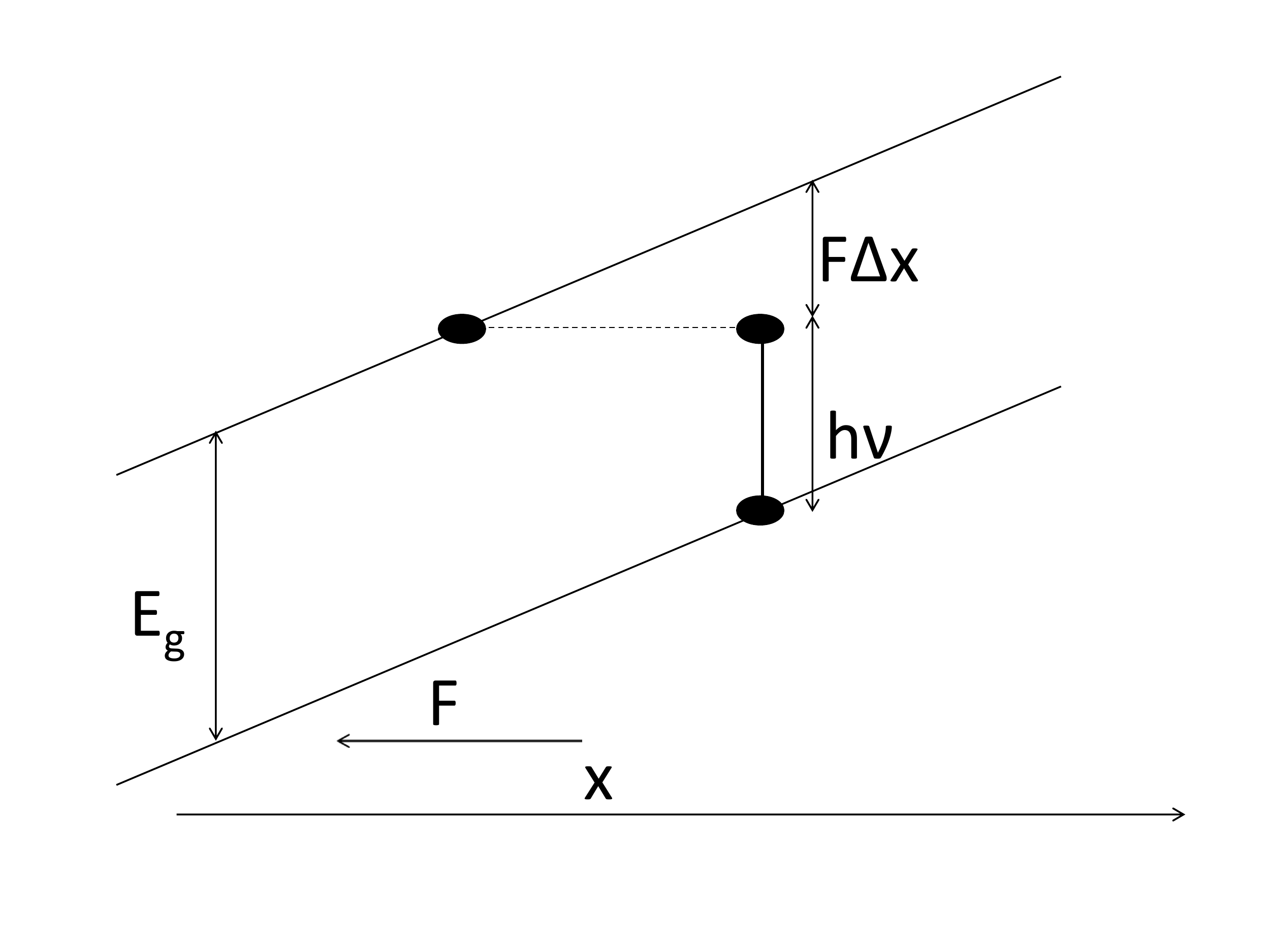}
\caption{Photon-assisted tunneling of an electron from the conduction to the valence band.}
\label{fig:photon}
\end{figure}

Another situation where common path interference effects may be important is photon assisted tunneling. In the absence of an electric field, the optical absorption coefficient goes to zero when photons of the incident light have energy below the band-gap. However, in the presence of an electric field, absorption of photons can take place as shown in Fig. \ref{fig:photon}. This is known as the Franz-Keldysh effect, or photon-assisted tunneling \cite{narrowgap}. If the bands are simply parabolic, the absorption coefficient when photon energy $\hbar\omega<E_g$ is proportional to $\exp(-f\left((E_g-\hbar\omega)/\Delta_0\right)^{3/2})$, where $f$ is a dimensionless constant. When the bands are non-parabolic it is easy to see that oscillations should be present in the optical absorption below the threshold energy as well. This may provide another experimental test of the common path interference phenomenon.

Thus, we have established that the common path interference identified in \cite{rahul} is in fact a universal phenomenon, arising whenever the low energy dispersion relation contains higher powers of $p^2$ (but not odd powers of $p$). The phenomenon is easiest to detect when the quadratic piece in the dispersion relation is absent (or comes with small co-efficient), however, the phenomenon is always present. It may be detected in transport experiments, or in optical absorption experiments designed to probe photon-assisted tunneling. Bilayer (and ABC multilayer) graphene are identified as particularly promising materials for detection of these effects, but other narrow gap materials such as several recently discovered topological insulators may also offer a playground for exploration of these phenomena. 



Acknowledgement: This research was supported by the Department of Energy, Office of Basic Energy Sciences; under grant DE-SC20002140 (S.J. and R.N.B.) and DE-FG02-ER45118 (E.J.M.). R.N. is supported by a PCTS fellowship.

\end{document}